\documentclass[12pt]{article}
\usepackage{epsfig,graphicx,amssymb,amsfonts}

\makeatletter \@addtoreset{equation}{section}

\makeatother

\def \be {\begin{equation}}
\def \ee {\end{equation}}
\def \bea {\begin{eqnarray}}
\def \eea {\end{eqnarray}}
\def \nn {\nonumber}

\def \R {{\textsf{I}\kern-.10em \textsf{R}}}
\def \T {{\textsf{T}\kern-.45em \textsf{T}}}
\def \C {{\textsf{C}\kern-.37em \textsf{C}}}
\def \Z {{\textsf{Z}\kern-.35em \textsf{Z}}}
\def \H {{\textsf{I}\kern-.10em \textsf{H}}}
\def \S {{\textsf{S}\kern-.37em \textsf{S}}}

\def \dels {\partial\kern-.5em / \kern.5em}
\def \As {{A\kern-.5em / \kern.5em}}
\def \Ds {D\kern-.7em / \kern.5em}

\newcommand{\ena}{\end{eqnarray}}

\def\bbox{{\,\lower0.9pt\vbox{\hrule \hbox{\vrule height 0.2 cm
\hskip 0.2 cm \vrule height 0.2 cm}\hrule}\,}}
\newcommand{\dsl}{\pa \kern-0.5em /}

\newcommand{\pa}{\partial}

\def \K {{\tt I\kern-.25em K}}

\begin{document}


\begin{titlepage}
\begin{center}

\hfill\parbox{4cm}{
{\normalsize\tt hep-th/0403094}} \\

\vskip .7in

{\LARGE \bf Twisting S-branes}

\vskip 0.5in

{\large John E.\ Wang} $\footnote{{\tt
hllywd2@feynman.harvard.edu}}$

\vskip 0.15in

{\it Department of Physics, Harvard University, Cambridge,
MA 02138, USA }\\
{\it Department of Physics, National Taiwan University
Taipei 106, Taiwan}\\

\end{center}

\vskip .5in

\begin{abstract}
\normalsize\noindent  Smooth time dependent supergravity solutions
corresponding to analytic continuations of Kerr black holes are
constructed and limits with a local de Sitter phase are found.
These solutions are non-singular due to a helical twist in space
and a fine tuning of the energy flow in the spacetime.  For the
extremal limit in which the mass and twist parameters are equal
the S-brane undergoes de Sitter expansion.  Subextremal limits
show the formation and decay of a twisted circle and closed
string tachyon condensation backreaction effects can be followed.
For small values of the twist deformation, a short lived
ergosphere envelopes the S-brane and leads to the production of
closed timelike curves.

\end{abstract}

\vfill

\end{titlepage}
\setcounter{footnote}{0}

\pagebreak
\renewcommand{\thepage}{\arabic{page}}
{\baselineskip=5mm\tableofcontents}


\section{Introduction}

Time dependent tachyon condensation \cite{stro,
roll,sugraSbranes,followS} has received considerable attention
recently due to exciting experimental results regarding the
accelerating nature of our universe. Attempts to find a source
for de Sitter space in the same way D-branes have near horizon
anti de Sitter space geometries, have included the introduction
of Spacelike branes \cite{stro} which describe the formation and
decay of an unstable brane.  While the formation and decay of an
unstable brane might be expected to be a smooth process, the
supergravity S-brane solutions of Refs.~\cite{stro, sugraSbranes}
contain singularities. It was not clear if these singularities
were a deficiency of the supergravity approximation or arose from
other factors, see Ref.~\cite{followS} for further details and
discussions on related issues.

Gutperle and Strominger proposed several resolutions of the
singularity problem including deforming the S-brane R-symmetry to
obtain smooth spacetimes. The R-symmetry corresponds to the
symmetry transverse to the S-brane worldvolume so reducing the
R-symmetry could localize the S-brane in space as well as time.
These deformed solutions would represent finely tuned incoming
energy forming a localized and unstable object which then decays
into outgoing radiation.  Non-singular S-brane configurations
have also recently been discussed in Ref.~\cite{nonsing}.

\begin{figure}[htb]
\begin{center}
\epsfxsize=6in\leavevmode\epsfbox{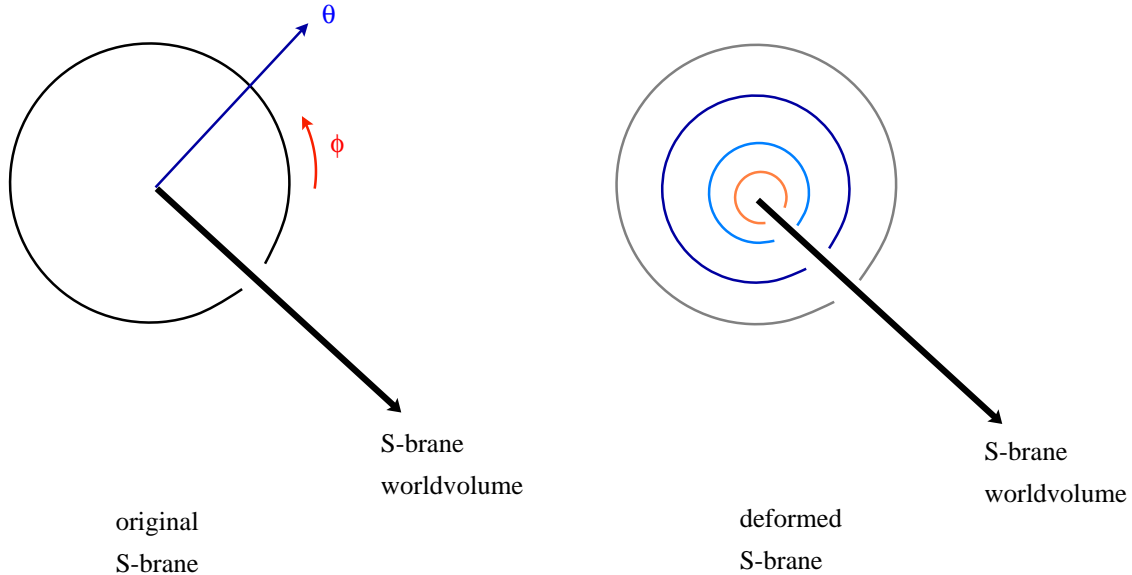} \caption{Due to
the original R-symmetry, the directions transverse to S-brane
worldvolume uniformly evolve in time.  By deforming the
R-symmetry it is possible to tune the energy flow in the
transverse directions.} \label{Sbrane-collapse}
\end{center}
\end{figure}

Inspired by the discussion of how to resolve S-brane singularities
in Ref.~\cite{stro} in this paper we begin an analysis of a class
of solutions with reduced R-symmetry. These solutions are analytic
continuations of rotating Kerr black holes and are free from
curvature singularities.   To distinguish these new solutions we
note that the original S0-brane in four dimensions was a
spacetime of the form

\be R^{1,1} \times H^2  \ee

\noindent and so had SO(1,2) R-symmetry.  The hyperbolic space
$H^2$ was transverse to the one dimensional worldvolume of the
S0-brane. For these new twisted S-branes the spacetime manifold
$\mathcal{M}$ is a fiber bundle over $R^{1,1}$ with fiber $H^2$
and so the hyperbolic space is non-trivially fibered over the
S-brane worldvolume

\be R^{1,1} \ltimes H^2 \ee

\noindent which reduces the R-symmetry to SO(2). At a fixed
radial coordinate $r$ the Kerr solution depends on $\theta$,
while at a fixed moment in time $\tau$ the S-brane has non-trivial
dependence in the radial $\theta$ direction as shown in
Figure~\ref{Sbrane-collapse}. Energy flows towards the S-brane
worldvolume, but the singularity is avoided because the regions
farther from the origin move more slowly.

Although the time and space coordinates of the metric are
effectively interchanged under Wick rotation, the spacetime
structure is not described by a ninety degree rotation of the
usual Kerr black hole Penrose diagram which is conventionally
drawn for the equatorial plane $\theta=\pi/2$.  This S-brane is
smooth and free from curvature singularities due to the
hyperbolic nature of the spacetime.  For the Kerr solution
$\theta$ is periodic so there is a special value at which to put
the ring singularity. Under analytic continuation, there is no
equivalent location to put the ring singularity and it therefore
vanishes.

These solutions are classified by a twist parameter, $a$, which
characterizes the R-symmetry breaking, and what we shall call a
mass parameter, $M$.  The extremal $a=M$, subextremal $a>M$ and
superextremal $a<M$ cases have distinct features and a general
discussion of their properties is given in Sec.~2.  Although the
Kerr solution has closed timelike curves near the ring
singularity, these twisted S-branes are free from closed timelike
curves for large twists $a\geq M$. In fact for $a>M$ there are no
horizons, no ergospheres and no curvature singularities.
Section~3 contains a discussion of the extremal limit which is
distinguished by having a localized two dimensional de Sitter
space limit, $dS_2\ltimes H^2$.  In Section~4 the subextremal case
is shown to be useful in the study of closed string tachyon
condensation backreaction on spacetime. The superextermal case is
discussed in Sec.~5 and is found to produce closed timelike
curves and an ergosphere.

\section{S-branes with reduced R-symmetry}

\subsection{Review of Sp-brane with SO(1,D-p-2) R-symmetry}

In four dimensions the S0-brane metric is

\be ds^2_{S0} = (1-\frac{2M}{\tau}- \frac{e^2}{\tau^2}) dz^2 -
\frac{d\tau^2}{1-\frac{2M}{\tau}- \frac{e^2}{\tau^2}} + \tau^2
(d\theta^2 + \sinh^2\theta d\phi^2)  \label{originalS} \ee

\noindent  where the S-brane worldvolume lies along the $z$
direction and $\theta,\phi$ are transverse directions. Time
symmetric S-branes are obtained by setting $M=0$. This solution
is the analytic continuation of Reissner Nordstrom black holes

\be \tau\rightarrow i r,  \  z\rightarrow i t, \ \theta
\rightarrow i \theta, \ M\rightarrow i M \ .\ee

\noindent The Penrose diagrams for these solutions are ninety
degree rotations of the Reissner Nordstrom Penrose diagrams, with
the difference being that every point on the Penrose diagram
represents a two dimensional hyperbola instead of a two sphere;
the original $SO(3)$ rotational symmetry becomes the $SO(1,2)$
R-symmetry after the analytic continuation.  In general starting
from a p-brane in D dimensions, the $SO(D-p-1)$ rotational
symmetry transverse to the brane worldvolume becomes the
$SO(1,D-p-2)$ Sp-brane R-symmetry.

In addition to the $M>e$ charged black holes there are the $M<e$
solutions with a naked timelike singularity.  The ninety degree
rotation of the naked singularity Penrose diagram has a
cosmological nature and is equivalent to the FRW Penrose diagram.
Cosmological implications of such S-brane solutions were discussed
in Refs.~\cite{Sbranecosmo} with the two cases, $M>e$ and $M<e$,
corresponding to two universes due to the curvature singularity.
Depending on the mass to charge ratio, there is either a big
bang/crunch or a timelike singularity.

\subsection{Twisting S-branes}

S-branes originally were homogenous and purely time dependent
constructions.  Their supergravity solutions were analytic
continuations of supergravity solutions of p-branes which were
isotropic in the transverse directions. Deforming S-branes and
localizing them in space is interesting and would require a
solution dependent on two variables, time and a radial distance
from the S-brane worldvolume. Solutions which depend on two
variables are unfortunately generally difficult to obtain. Given
the fact though that the S0-brane is the analytic continuation of
a black hole solution it is possible to ask if there are any known
non-isotropic black holes.  The most general solution with mass,
charge and angular momentum in four dimensions is the rotating
charged black hole which is the focus of this paper.

The electrically charged Kerr-Newman metric in Boyer-Lindquist
coordinates is

\begin{eqnarray}
ds^2_{Kerr} & = &-(\frac{\Delta - a^2 \textup{sin}^2
\theta}{\rho^2}) dt^2 - \frac{2 a \textup{sin}^2 \theta (r^2 +
a^2 - \Delta)}{\rho^2} dt \ d\phi \\
&& + [ \frac{(r^2 + a^2)^2 - \Delta a^2 \textup{sin}^2
\theta}{\rho^2} ] \textup{sin}^2 \theta d\phi^2 +
\frac{\rho^2}{\Delta} dr^2 + \rho^2 d\theta^2  \nonumber \\
A & = & - \frac{e r}{\rho^2} ( dt- a \textup{sin}^2 \theta \
d\phi)
\end{eqnarray}
\begin{equation}
\rho^2 = r^2 + a^2 \textup{cos}^2 \theta \hspace{.3in} \Delta =
r^2 + a^2 + e^2 - 2Mr \ .
\end{equation}

\noindent Here $M,e,a$ are the mass, charge and angular momentum
parameters while $A$ is the vector potential.  The interesting
features of this solution persist in the absence of charges so we
will set the electric (and magnetic) charges to zero.   One may
consider the angular momentum parameter to be a different charge
carried by the black hole.  The neutral Kerr metric is

\be ds^2_{Kerr} = -\frac{\Delta}{\rho^2}(dt - a \sin^2\theta
d\phi)^2 +\frac{\sin^2\theta}{\rho^2}[(r^2+a^2) d\phi - a dt]^2
+\frac{\rho^2}{\Delta}dr^2 +\rho^2 d\theta^2  \ee

\noindent which has been written in a form convenient for
checking various properties.  When $a<M$, the Kerr black hole has
two horizons and a ring singularity.  For the extremal case
$a=M$, the horizons become degenerate and when $a>M$ the
singularity is naked.

\begin{figure}[htb]
\begin{center}
\epsfxsize=5in\leavevmode\epsfbox{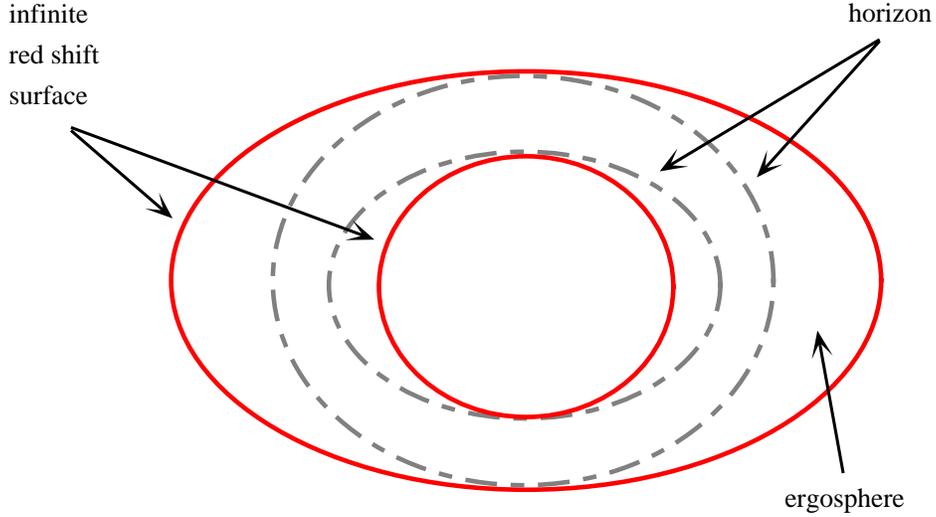}
\caption{For $a<M$ the Kerr black hole has two horizons and an
ergosphere.} \label{Kerrpicture}
\end{center}
\end{figure}

To obtain the twisted S-brane metric we analytic continue three
coordinates along with the angular momentum and mass parameters
\begin{equation}
t \ \mapsto \ i z, \hspace{.3in} \theta \ \mapsto \ i \theta,
\hspace{.3in} r \ \mapsto \ i \tau;  \hspace{.3in} a \
\rightarrow \ i a , \hspace{.3in} M \ \rightarrow \ i M
\end{equation}

\noindent where our convention is not to introduce minus signs in
the analytic continuation.   The Wick rotation has the following
effect
\begin{equation}
\rho^2 \ \rightarrow \ -\tau^2 - a^2 \cosh^2\theta = -\rho^2,
\hspace{.3in} \Delta \ \rightarrow \ -\tau^2 + 2M\tau - a^2 =
-\Delta \ .
\end{equation}

\noindent The twisted S0-brane metric we will focus on is

\begin{eqnarray} ds^2_{twisted \ S0}& = & \frac{\Delta + a^2
\sinh^2\theta}{\rho^2} dz^2 + \frac{2a
\sinh^2\theta(\tau^2+a^2-\Delta)}{\rho^2}dz d\phi   \label{twistedS} \\
& & + [ \frac{(\tau^2+a^2)^2 + \Delta a^2 \sinh^2\theta}{\rho^2}]
\sinh^2\theta d\phi^2 + \rho^2(-\frac{d\tau^2}{\Delta} +
d\theta^2) \nonumber \\
& =& -\frac{\rho^2}{\Delta}d\tau^2 +\rho^2 d\theta^2 +
\frac{\Delta}{\rho^2}(dz - a \sinh^2\theta d\phi)^2
+\frac{\sinh^2\theta}{\rho^2}[(\tau^2+a^2) d\phi + a dz]^2
\nonumber
\end{eqnarray}
\be \rho^2 = \tau^2 + a^2 \cosh^2\theta , \hspace{.3in}
\Delta=\tau^2+a^2-2M\tau \label{functions} \ee

\noindent where we grouped the terms to hi-light the dependence on
the factor $\Delta$.

\subsection{Absence of curvature singularities and closed timelike curves}

Several properties of this spacetime may be inferred from our
knowledge of Kerr black holes.  When the parameter $a=0$ this
gives the original S0-brane of Eq.~\ref{originalS}, so this is a
new twisted deformation of known S-branes. For large values of
$\tau$, $\tau$ acts as the time coordinate, although for various
parameter ranges there are horizons where $\Delta$ changes sign
and so $\tau$ is not necessarily always timelike. The $\theta$
dependence means that the metric can represent a spatially
localized source and due to the metric cross term $g_{z\phi}$
there is a coupling of two directions $z$ and $\phi$.

For large values of angular momentum
\begin{equation}
a^2 > m^2 \ \rightarrow \ \Delta >0 \label{largemom}
\end{equation}
the Kerr solution does not have a horizon, where $\Delta=0$.  It
is easy to check that after Wick rotation Eq.~\ref{largemom} still
holds, so the twisted S-brane does not have horizons for large
values of the parameter $a$. The Kerr solution is also known to
have a ring curvature singularity at $\rho^2=0$
\begin{equation}
\rho^2 = 0 \ \rightarrow \  r=0, \ \theta=\frac{\pi}{2} \ .
\end{equation}
\noindent The curvature for the S-brane is similarly dependent on
\begin{equation}
\rho^2 = \tau^2 + a^2 \cosh^2\theta \ .
\end{equation}
The difference however is that while $\tau$ can be zero, there are
no zeros for hyperbolic cosine so $\rho \neq 0$ for real values of
$\theta$. Therefore the hyperbolic nature of the transverse space
guarantees that this S-brane has no curvature singularities.
Intuitively the reason that the singularity disappears is due to
the fact that there is a special and unique value of $\theta$ at
which to put the ring singularity in the case of the rotating
black hole, but there is no equivalent value of $\theta$ in
hyperbolic space to put a singularity.  The conformal structure
of this new spacetime is very different from the original
S-branes and Kerr spacetimes. As shown by this example, however,
reducing the R-symmetry is one way to cure the S-brane
singularity problem.

Although for real values of $\theta$ the spacetime is smooth, it
might be interesting to examine the fact that there is a
singularity for $\theta=-i\pi/2$.  Analogous observations have
been made about Sen's rolling tachyon solution which can be
thought of as a periodic array of D-branes in imaginary time
\cite{imaginary}. For this S-brane however the singularities lie
on a circle at an imaginary $\theta$ value.  It would be
interesting to clarify this point and understand if the
singularity in complex directions corresponds to a specific
tachyonic mode instability.

A known problem with the Kerr solution is the existence of closed
timelike curves. The problem is that the periodic angular
coordinate $\phi$ changes from being timelike to spacelike since
the coefficient

\be g_{\phi \phi}=(\frac{(r^2+a^2)^2 - a^2 \Delta
\sin^2\theta}{r^2 + a^2 \cos^2\theta}  ) \sin^2\theta \ee

\noindent changes sign from positive to negative values for

\be (r^2 +a^2)^2 < a^2 \sin^2\theta (r^2 + a^2 - 2Mr) \ . \ee

\noindent Near the ring singularity at $\theta=\pi/2$, and for $r$
small and negative, the above condition is satisfied and we have
closed timelike curves.

It is then imperative to check if this twisted S-brane also has
closed timelike curves.  One difference is that the time
direction is now along $\tau$ and

\be g_{\tau\tau} = -\frac{\rho^2}{\Delta} \ee

\noindent which is governed by the sign of $\Delta$, which we have
said is always positive for $a>M$.  When the twist parameter $a$
is large relative to $M$, then $z$ and $\phi$ coordinates are
always spacelike coordinates.  It is also possible to explicitly
check that the condition $g_{\phi\phi}<0$ implies

\be (\tau^2 + a^2)^2 < -a^2 \sinh^2\theta (\tau^2 +a^2 - 2M\tau)
\ee

\noindent which is satisfied if and only if

\be \tau^2 + a^2 - 2M\tau < 0 \ . \ee

\noindent For $a\geq M$ there are no closed timelike curves and
the spacetime does not rotate. When the angular momentum
parameter is small, $a<M$, the S-brane does have an ergosphere and
closed timelike curves as discussed in Sec.~5.

Adding charge does not significantly change the properties of
this S-brane.  For electrically charged S-branes we find that the
solution is still smooth and free from closed timelike curves as
long as $a^2-e^2\geq M^2$.

\subsection{General properties}

The Kerr solution has two null vectors

\begin{eqnarray} l^\mu & = & \frac{r^2 + a^2}{\Delta} (\frac{\partial}{\partial
t})^\mu + \frac{a}{\Delta} (\frac{\partial}{\partial \phi})^\mu +
(\frac{\partial}{\partial r})^\mu \\
n^\mu &=& \frac{r^2 + a^2}{2 \rho^2} (\frac{\partial}{\partial
t})^\mu  + \frac{a}{2 \rho^2} (\frac{\partial}{\partial \phi})^\mu
- \frac{\Delta}{2 \rho^2} (\frac{\partial}{\partial r})^\mu
\end{eqnarray}

\noindent and likewise the twisted S-brane has two null vectors

\begin{eqnarray}
\tilde{l}^\mu & = & \frac{\tau^2 + a^2}{\Delta}
(\frac{\partial}{\partial z})^\mu + \frac{a}{\Delta}
(\frac{\partial}{\partial \phi})^\mu + (\frac{\partial}{\partial
\tau})^\mu \\
\tilde{n}^\mu & = & -\frac{\tau^2 + a^2}{2 \rho^2}
(\frac{\partial}{\partial z})^\mu  - \frac{a}{2 \rho^2}
(\frac{\partial}{\partial \phi})^\mu + \frac{\Delta}{2 \rho^2}
(\frac{\partial}{\partial \tau})^\mu
\end{eqnarray}
\be \tilde{l}^2=0=\tilde{n}^2, \hspace{.3in} \tilde{l}\cdot
\tilde{n}=-1  \ . \ee

\noindent From these two null vectors it is possible to construct
a Killing two tensor which is useful for simplifying calculations
and checking properties of this S-brane.   The S-brane metric is
manifestly independent of $z$ and $\phi$ so there are also two
Killing vectors, $\partial/\partial z$ and $\partial/\partial
\phi$. An interesting point is that near $\Delta=0$, the null
vectors are also Killing vectors.

Although the twisted Sbrane metric is independent of the
coordinate $z$, there is a metric cross term $g_{z\phi}$.
Concretely the metric is invariant only under the combined
discrete symmetry

\be z \ \rightarrow \ -z , \hspace{.3in} \ \phi \ \rightarrow
-\phi \ee

\noindent demonstrating that there is a twist in space, which is
a spiral or helix structure along these two directions.  The cross
term in the metric in these two spatial directions is reminiscent
of twisted circles discussed in Ref.~\cite{Dowker}. Interestingly
it is shown in Sec.~4 that while supergravity S-branes were
originally conceived to have implications for open string tachyon
condensation, this twisted S-brane has implications for closed
string tachyon condensation \cite{closedtachyon} in particular
the time dependent formation and decay of a twisted circle.

The horizons of the Kerr black hole are defined to be where
$g_{rr}=\infty$, so $\Delta=0$, and they occur at

\be r_{horizon} = M \pm \sqrt{M^2 -a^2}  \ . \ee

\noindent There are also infinite red shift surfaces where
$g_{tt}=0$ at

\be   r_{ergosphere}= M \pm \sqrt{M^2 - a^2 \cos^2\theta} \ee

\noindent which are the boundaries of the ergosphere and which
begin farther out from the origin than the horizon

\be r_{ergo+} > r_{horizon+} \ . \ee

For the S-branes the situation is similar but when $a>M$,
$\Delta\neq 0$, the metric is smooth and there are no metric
degeneracies.  If $a\leq M$ then $g_{\tau\tau}$ does not go to
zero but can go to infinity. Instead of an infinite red shifting
surface there is a infinite blue shifting surface. There is also
a surface on which $g_{zz}=0$ which we regard as the boundary of
an ergosphere.  These points are further discussed in Sec.~5.

\subsection{Far from origin and late time behavior}

It is interesting to examine properties of the twisted S-brane
far from the origin. In the limit $\cosh^2\theta >> \tau^2/a^2$
the metric is

\be ds^2 =  e^{2\theta} [ -\frac{a^2 d\tau^2}{\tau^2+a^2-2M\tau} +
(\tau^2+a^2-2M\tau) d\phi^2  + a^2 d\theta^2] + dz^2 +
\frac{\tau^2+a^2-\Delta}{a} dz d\phi \ . \label{faraway} \ee

\noindent At large values of $\theta$, the effect of the cross
term is small and can be neglected.   If we also take the limit
$\tau>> M, a$ the metric is

\be ds^2 = e^{2\theta} [ -\frac{a^2 d\tau^2}{\tau^2} + \tau^2
d\phi^2  + a^2 d\theta^2] + dz^2  \ee

\noindent which we further rewrite so that $\theta$ acts as a
radial coordinate

\be R=a e^{\theta}, \hspace{.3in}  T= \ln \frac{\tau}{\tau_0} \ee
\begin{eqnarray}
ds^2 & =& dR^2 - \frac{R^2}{\Delta} d\tau^2  +
\frac{\Delta}{a^2} R^2 d\phi^2 + dz^2 \\
& =& e^{2\theta} [ -dT^2 + \tau_0^2 e^{2T} d\phi^2 + a^2
d\theta^2]
+ dz^2 \nonumber   \\
&= & dR^2- R^2 dT^2 + \frac{\tau_0^2}{a^2} R^2 e^{2T} d\phi^2 +
dz^2 \ .
 \nonumber
\end{eqnarray}

\noindent The metric in this limit is two dimensional Rindler
space with exponential expansion in the angular direction which
will have the interpretation of being the result of a passing
pulse of energy.

Next examine Eq.~\ref{faraway} for values of $\tau$ which
minimize $\Delta$. The physical significance of this time is that
it will serve as an interesting initial condition for the time
evolution of this solution.   For the case $a>M$, the metric near
$\tau=M$ is a Rindler wedge times a line of flat cones

\be ds^2_{far \ cone} = dR^2 - R^2 d\tau^2 + dz^2 +
\frac{\Delta}{a^2}R^2 d\phi^2  \ . \label{cone} \ee

\noindent Returning to the usual Minkowski coordinates the above
limit gives the metric

\be ds^2_{far \ cone} = -dt^2 +dx^2 +dz^2 + \frac{\Delta}{a^2}
(x^2-t^2) d\phi^2 \ . \ee


Examining the late time limit where $\tau^2/a^2 >> \cosh^2\theta$
we have

\begin{equation}
ds^2 = -d\tau^2 + \tau^2 dH_2^2 + dz^2 -\frac{2M}{\tau}(a
\sinh^2\theta d\phi + dz)^2
\end{equation}

\noindent a Milne wedge with a twist whose effect disappears at
late times.  For observers in the past and future light cones, the
R-symmetry breaking is therefore spontaneous in time.

\section{Extremal limit $a=M$}

One of the original motivations for introducing Spacelike Branes
was to find an object in string theory which could source de
Sitter space.  In fact there are E-branes \cite{Hull} which do
have a de Sitter space near horizon limit.  These are obtained by
starting from a BPS brane, and analytically continuing both the
electric charge and the mass.  Although this does lead to a de
Sitter space limit, analytically continuing the charge leads to
type II* string theories with a wrong sign kinetic energy term
for the field strength.  Known charged S-branes in type II string
theories do not have such an extremal limit.  To find an extremal
limit related to de Sitter space and also stay in type II string
theories, one apparently needs a new type of charge.

The twisted S-branes in this paper are solutions to the vacuum
Einstein equations but can be interpreted as having a charge
arising from twisting and not from a field strength and so they
should exist in the usual string theories. It is known that Kerr
black holes do have an interesting extremal limit, $a=M$.  We
begin by examining this extremal S-brane limit and find that the
S-brane worldvolume undergoes an expanding de Sitter phase.

\subsection{Review of extremal Reissner Nordstrom horizon}

The extremal Reissner Nordstrom black hole metric

\be ds^2_{RN}= -(1-\frac{2M}{r} +\frac{M^2}{r^2})dt^2 +
\frac{dr^2}{1-\frac{2M}{r}+\frac{M^2}{r^2}} + r^2 dS^2_2 \ee

\noindent has a near horizon limit

\be r= M + \delta \ee

\be ds^2_{RN} = -\frac{\delta^2}{M^2} dt^2 + \frac{M^2
d\delta^2}{\delta^2} + M^2 dS^2_2  \ee

\noindent which is $AdS_2\times S_2$.  Starting from this
solution it is possible to obtain a near horizon limit which is de
Sitter space instead of anti de Sitter space, if the mass and the
charge are Wick rotated $e=M\rightarrow iM$.  These are solutions
however in type II* theories with wrong sign kinetic energy terms
since the field strength contribution to the action $F^2\sim e^2$
becomes $-F^2$.

\subsection{Near horizon limit}

We begin by examining the terms in the S-brane metric
Eq.~\ref{twistedS} along the $z,\phi$ directions

\be \frac{\Delta}{\rho^2} dz^2 + \frac{2 \Delta a
\sinh^2\theta}{\rho^2} dz d\phi + \frac{a^2 \Delta
\sinh^4\theta}{\rho^2} d\phi^2 \label{crossterms} \ee

and

\be \frac{a^2 \sinh^2\theta}{\rho^2} dz^2 + \frac{2 a
\sinh^2\theta (\tau^2+a^2) }{\rho^2} dz d\phi + \frac{
(\tau^2+a^2)^2 \sinh^2\theta}{\rho^2} d\phi^2 \ . \ee

\noindent Given that this S-brane is localized in space as well as
time, we look at times which minimize $\Delta$ and also approach
the S-brane worldvolume so $\theta$ is small

\be \tau = M +\delta ,\hspace{.3in}  0\leq \Delta  =
(\tau-M)^2=\delta^2 << |\tau^2 +a^2| , \hspace{.35in} |\Delta a^2
\sinh^2\theta | << (\tau^2 +a^2)^2 \ . \label{nearSbrane} \ee

\noindent In this limit the second and third terms in
Eq.~\ref{crossterms} can be neglected and the metric is

\begin{eqnarray} ds^2_{close \ extremal}& =& -\frac{\rho^2}{\Delta} d\tau^2 + \rho^2 d\theta^2 +
\frac{\Delta}{\rho^2} dz^2 + 2a^2 \sinh\theta^2 ( d\phi +
\frac{1}{2a} dz)^2 \\
& = & -\frac{ 2M^2 d\delta^2}{\delta^2} + \frac{\delta^2}{2M^2}
dz^2 + 2M^2 [ d\theta^2 + \sinh^2\theta ( d\phi + \frac{dz}{2M})^2
]
\end{eqnarray}

\noindent a twisted hyperbolic space fibered non-trivially over
two dimensional de Sitter space

\be dS_2 \ltimes H^2 \ . \ee

\noindent Writing the metric in its exponential form

\be T= \sqrt{2} M \ln \frac{\delta}{\delta_0} , \hspace{.3in}
\delta = \delta_0 e^{\frac{T}{\sqrt{2}M}} \ee
\be ds^2_{close \ extremal} = -dT^2 + \frac{\delta_0^2}{2M^2}
e^{\frac{T}{\sqrt{2}M}} dz^2 + 2M^2[ d\theta^2 + \sinh^2\theta (
d\phi - \frac{1}{2M}dz )^2] \label{exponential} \ee

\noindent one finds that the Hubble parameter is
$H^{-1}=\sqrt{2}M$.  After a time $\tau\sim2M$ the near S-brane
limit of Eq.~\ref{nearSbrane} is no longer valid; due to the
instability of this configuration, it eventually disperses.  This
causes the exponential expansion phase to slow down and the
spacetime approaches flat space.  If we add electric or magnetic
charge to this S-brane, it is still possible to find an extremal
limit similar to Eq.~\ref{exponential}, but the relative metric
coefficients will be changed.

To remove the twist near the origin one may use the twisted
coordinate with non-standard periodicity

\be \tilde{\phi} \rightarrow \phi - \frac{z}{2M} \ee

\noindent to obtain a local region of $dS_2\times H^2$.  If we
further periodically identify the $z$ direction, this solution
apparently demonstrates how flat spacetime decays to a Melvin
universe in the far future. It would be interesting to embed this
solution in string theory and see what role it plays in the
proposed dualities and relations \cite{zeroduality} between type
0 and type II string theories.

While Wick rotating the electric charge leads to solutions in
*-theories, Wick rotating the angular momentum parameter leads to
a twist parameter.  Instead of having S-branes charged under
electro-magnetic fields, we deform the spacetime with a twist to
obtain a region of de Sitter space near the S-brane worldvolume.

\begin{figure}[htb]
\begin{center}
\epsfxsize=5in\leavevmode\epsfbox{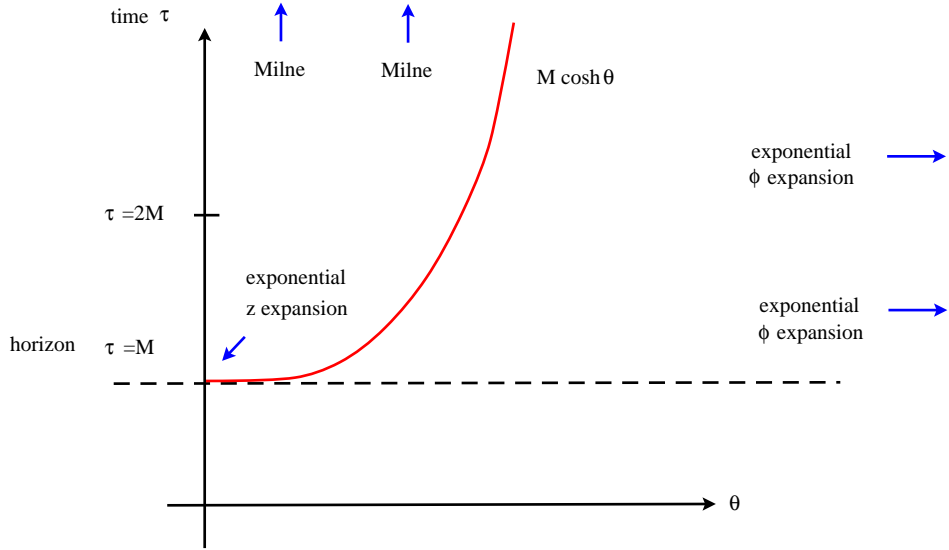}
\caption{The neighborhood of the S-brane worldvolume undergoes a
de Sitter expansion phase for times $\tau\sim M$. The expansion
slows down into a Milne spacetime after $\tau\sim 2M$. Far from
the origin, we have regions of exponential expansion of the $\phi$
direction.  The curve $\tau=M\cosh\theta$ marks the boundary
between the two types of behaviors.} \label{farfromlump-extremal}
\end{center}
\end{figure}

\section{Subextremal case $a>M$}

For the parameter range $a>M$ these subextremal S-branes have
neither horizons nor ergospheres, and the near S-brane limit
qualitatively differs from the extremal case.  The time evolution
of this solution is interpreted as showing the gravitational
backreaction of local closed string tachyon condensation.  In
particular the solution shows the decay of a cone whose tip is a
twisted circle.  As the twist unwinds, this causes the cone to
exponentially expand and open up into flat space.

\subsection{Review of twisted circles}

A twisted circle \cite{Dowker} is a non-standard identification
of $R^3$

\be ds^2= dr^2 + r^2 d\phi^2 +dz^2 \ee

\be \phi \sim \phi + 2\pi n_1 + 2\pi n_2 R B ,\hspace{.3in} z\sim
z+ 2\pi n_2 R \ee

\noindent where the unusual feature is that the $z$ direction is
coupled with the $\phi$ direction.  To recover a more standard set
of coordinates one can introduce a twisted coordinate

\be \tilde{\phi}=\phi - Bz \label{twistcoord} \ee

\noindent with standard periodicity $\tilde{\phi}\sim \tilde{\phi}
+ 2\pi n_1$. The metric then becomes

\be ds^2 = dr^2 +r^2(d\tilde{\phi} + B dz)^2 + dz^2 \ee

\noindent and under dimensional reduction of the $z$ direction,
the twist, which is characterized by the constant $B$, becomes a
magnetic field.

\subsection{Near extremal limit}

Although when $a>M$ there is no horizon, the metric simplifies
for $\tau=M$ since this minimizes $\Delta(\tau=M)= a^2-M^2$.  If
we also take the near extremal limit, $a\approx M$, and near
origin limits $\theta<<1$

\be \tau = M +\delta,  \ \  \Delta = a^2 -M^2 +\delta^2, \ \
\rho^2\sim 2a^2 \ee

\be \delta^2<<a^2-M^2=\epsilon, \ \ \ \ \delta<< a, \ M, \ \ \ \
\epsilon<< M \delta \ee

\be a \theta << \delta << \sqrt{\epsilon} << \sqrt{\delta M} <<
M,a \ee

\noindent the metric is

\be ds^2_{NE-} =-\frac{\rho^2}{\epsilon} d\delta^2 + \rho^2
d\theta^2 + \frac{\epsilon}{\rho^2} dz^2 +
\frac{\sinh^2\theta}{\rho^2}[\rho^2
d\phi + adz]^2 \ .
\ee

\noindent Rescaling the time and $z$ coordinates and taking the
limit $\sinh\theta\sim \theta$ the metric becomes

\be z'= \frac{\sqrt{\epsilon}}{\rho} z= \sqrt{\frac{\epsilon}{2}}
\frac{z}{M}, \hspace{.3in} \delta=\sqrt{\frac{\epsilon}{2}}
\frac{T}{M} \ee

\be ds^2_{NE-} = -dT^2 + dR^2 + R^2 [
d\phi - \frac{1}{\sqrt{2\epsilon}} dz']^2 + dz'^2 \ .
\label{NEplus} \ee

\begin{figure}[htb]
\begin{center}
\epsfxsize=4in\leavevmode\epsfbox{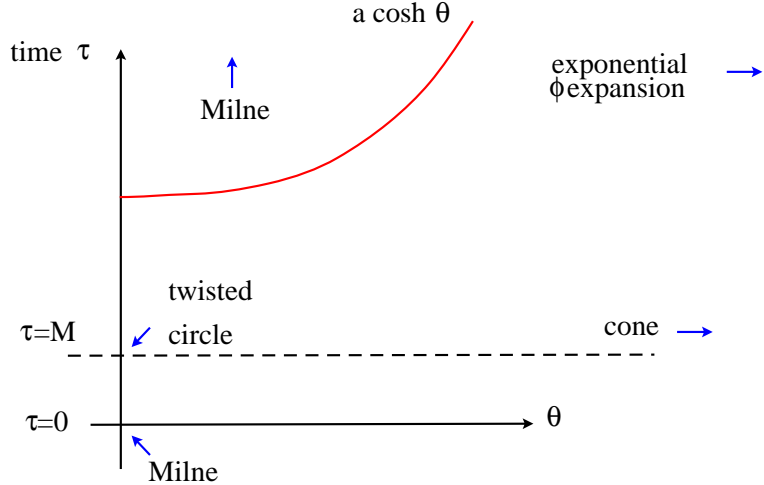} \caption{For
the subextremal case $a>M$, near the origin we have a twisted
circle at $\tau=M$ and which disperses at time $\tau=2M$ and
flattens out to Milne after $\tau=a$.  Far from the origin, there
are regions of Rindler near time $\tau=M$.  After a time a pulse
from the origin causes an exponential expansion in the angular
directions and after the pulse passes the space becomes flat. }
\label{farfromlump}
\end{center}
\end{figure}

\begin{figure}[htb]
\begin{center}
\epsfxsize=5in\leavevmode\epsfbox{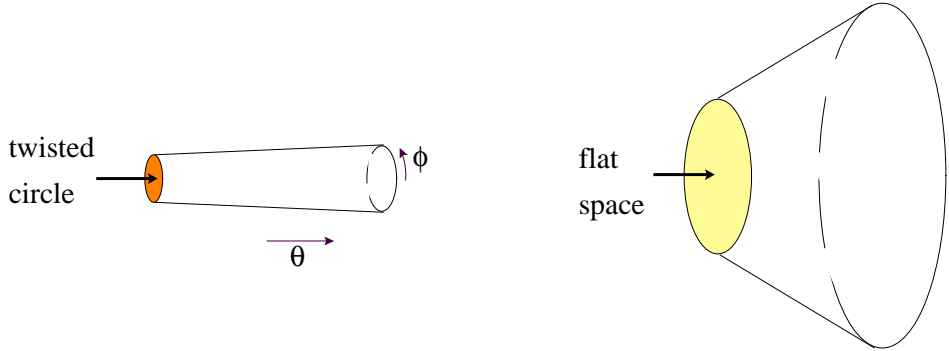} \caption{For
the near extremal solution, the tip of the cone is a twisted
circle at time $\tau=M$.  Going forward in time the circle
untwists and starting from the tip the cone opens up.}
\label{expandingcone}
\end{center}
\end{figure}

\noindent In this limit if we periodically identify the $z$
direction, we obtain a twisted circle. As time passes, there will
be an exponential expansion phase, and eventually the twist
unwinds leaving a growing region of flat space as shown in
Figs.~\ref{farfromlump},\ref{expandingcone}.  Recalling that at
$\tau=M$, the spacetime far from the origin is a cone (times a
line see Eq.~\ref{cone}), then we see that this solution
represents the time evolution of a cone whose tip is replaced by
a twisted circle.  The importance of $\tau=M$ is that it serves
as an interesting initial condition for time evolution for the
system.  Furthermore this solution is not symmetric in time so
there are actually two ways in which the solution unwinds.

\subsection{Compactifying the $z$ direction}

Although it does not seem necessary to periodically identify the
$z$ direction, it is natural so the metric has a twisted circle
limit.  In addition if we wish to have a sensible Kaluza-Klein
compactification near $\tau=M$, it is necessary that the magnetic
field be small relative to the compactification scale, $B<<1/R$.
From Eq.~\ref{NEplus} we see that the magnetic field is small
$B=(2\epsilon)^{-1/2}$ and so the compactification radius for the
$z$ direction must be smaller than $\epsilon^{1/2}$.

Under compactification this solution has the interpretation as
the formation and decay of a fluxbrane. Fluxbranes have been
shown to arise from a special scaling limit of a brane and
anti-brane pair \cite{fbranelimit}, so the parameter $a$ might be
interpreted as how much charge flows along the S-brane
worldvolume; the D-brane and anti D-brane annihilate via the
twist in the S-brane.

Under periodic identification of $z$ we do not introduce closed
timelike curves since we are in the limit $a>M$ where $\Delta$ is
always positive.  Compactifying in the extremal case $a=M$ we
would obtain closed null curves and it would be interesting to
discuss this in relation to work in Ref.~\cite{nullorb}. In
addition, under compactification the extremal case would lead to a
metric degeneracy at $\tau=M$ and $\theta=0$.  To observers bound
to the S-brane worldvolume this would appear as a big bang type
singularity which is smoothed out by opening a new spatial
dimension.

\begin{figure}[htb]
\begin{center}
\epsfxsize=3in\leavevmode\epsfbox{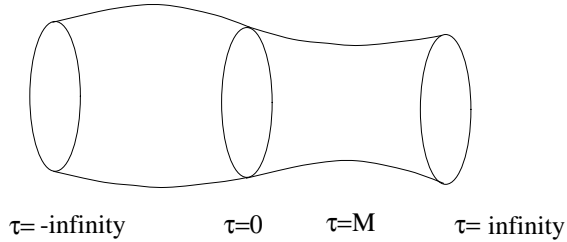} \caption{Size
of the compactified $z$ direction at fixed $\theta$.}
\label{zcircle-size}
\end{center}
\end{figure}

The metric for the circle direction

\be g_{zz}= 1- \frac{2M\tau}{\rho^2} \ee

\noindent starts with value one at past infinity, increases, then
decreases until we get to $\tau=0$ where it is one again. Then the
circle size is less than one, decreasing and going back to one at
future infinity.  At $\tau=M$ the circle reaches it minimum size
which can go to zero in the extremal limit.  As we go farther
from origin, the time dependence is less pronounced and delayed
relative to what occurs near the origin. See
figure~\ref{zcircle-size}.

Under the usual Kaluza-Klein reduction this spacetime has the
interpretation of a the time dependent formation and decay of a
fluxbrane.   If we use the twisted coordinates as in
Eq.~\ref{twistcoord} to remove the twist at $\tau=M$, an
interesting interpretation of this is that we are seeing how flat
space decays into a twisted circle at late times.  As in the
extremal case it will be interesting to embed this solution in
string theory.

\section{Superextremal case $a<M$}

In this section we examine $a<M$ superextremal S-branes which are
distinguished by their horizons, ergospheres and creation of
closed timelike curves.

\subsection{Ergosphere}

For small values of the twist parameter, $a<M$, one might assume
that these solutions become qualitatively similar to the original
S-brane solutions but in fact a small twist parameter drastically
distorts the spacetime structure.  These solutions have an
ergosphere which exists between two high blue shifting times.

For the Kerr black hole the ergosphere always lies outside the
horizon.  On the other hand S-branes have surfaces where $g_{\tau
\tau}=\infty$

\be \tau_{\pm} = M \pm \sqrt{M^2 - a^2} \ee

\noindent and the Killing horizons $g_{zz}=0$ occur when

\begin{equation}
\tau_{ergo}= M \pm \sqrt{M^2 - a^2 \cosh^2\theta }
\end{equation}

\noindent with

\be \tau_- \leq \tau_{ergo} \leq \tau_+ \ . \ee

\begin{figure}[htb]
\begin{center}
\epsfxsize=5in\leavevmode\epsfbox{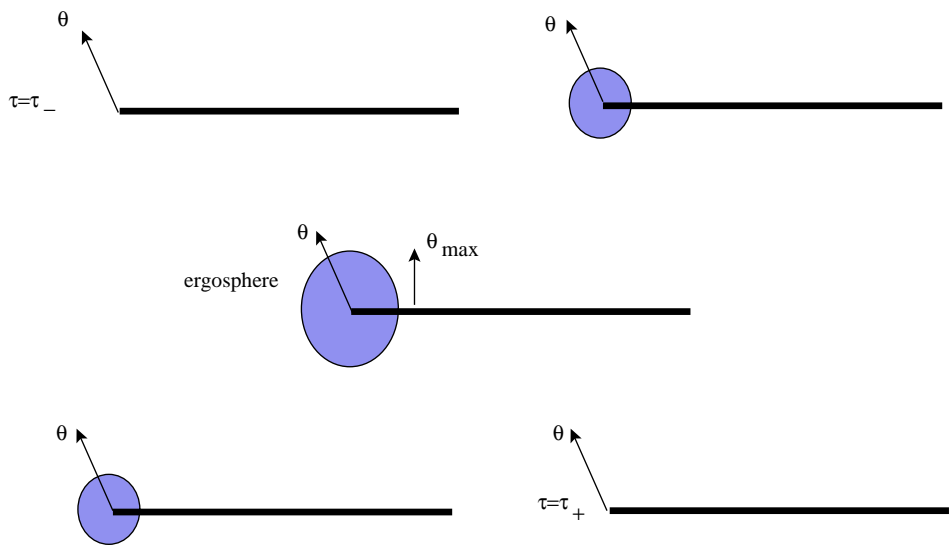}
\caption{There is an ergosphere between the times $\tau_-,\
\tau_+$. The ergosphere only exists in a neighborhood of the
S-brane and extends out to a maximum distance $\theta_{max}$. The
five figures are shown sequentially in time from left to right,
top to bottom.} \label{KerrSbrane-horizons-ergosphere}
\end{center}
\end{figure}

\noindent We will call the region between the Killing horizon and
the origin to be the S-brane's ergosphere.  The ergosphere extends
out to a maximum $\theta$ value for $\tau=M$ and then it
disappears as illustrated in
Figure~\ref{KerrSbrane-horizons-ergosphere}.  In some sense the
ergosphere is hidden.

Due to the change in spacetime signature this solution does
rotate for a period of time between $\tau_-$ and $\tau_+$.

\subsection{Near extremal limit}

For $a<M$ there are two non-degenerate roots of $\Delta$ and near
$\tau=\tau_+ +\delta$ the time component of the metric is

\be -\frac{\rho^2}{\Delta} d\tau^2 = -\frac{ \tau_+^2
+a^2}{\tau_{+-}\delta} d\delta^2, \hspace{.3in} \tau_{+-} =
\tau_{+} -\tau_{-} \ . \ee

\noindent Taking the same near S-brane limits as in
Eq.~\ref{nearSbrane} and using a coordinate transformation

\be  T = \frac{2\sqrt{2}\rho \delta^{1/2}}{\tau_{+-}^{1/2}},
\hspace{.3in} \delta = \frac{\tau_{+-}}{4\rho^2} T^2,
\hspace{.3in} z'=\frac{\tau_{+-} z}{2 \rho^2}= \frac{\tau_{+-}
z}{4M^2} \ee

\begin{eqnarray} ds^2_{NE+} &=& -dT^2 + \rho^2 d\theta^2 + \frac{\tau_{+-}
\delta}{\rho^2} dz^2 + \frac{\sinh^2\theta}{\rho^2} [ (\tau_+^2
+a^2
) d\phi + a dz]^2 \\
&=& -dT^2 +  T^2 dz'^2 + 2M^2\{ d\theta^2 + \sinh^2\theta
 [
d\phi +4 M\frac{dz'}{\tau_{+-}}]^2 \} \nonumber
\end{eqnarray}

\noindent the metric again describes a twisted circle for small
$\theta$ which then untwists.

\begin{figure}[htb]
\begin{center}
\epsfxsize=4in\leavevmode\epsfbox{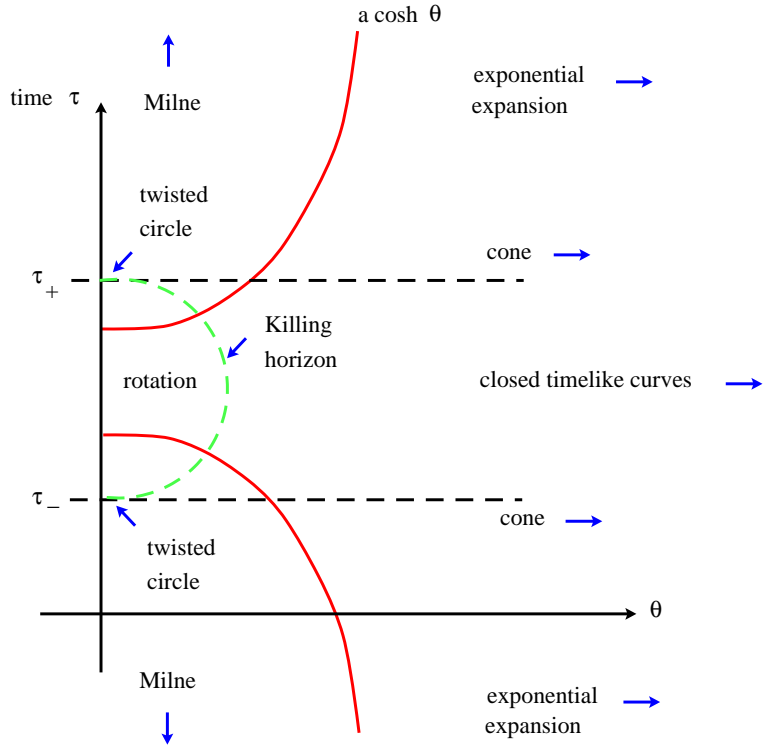}
\caption{For the superextermal case $a<M$ there are two horizons
and an ergosphere.  Near the origin we have a twisted circle.
Closed timelike curves exist for $\tau_-<\tau<\tau_+$.}
\label{farfromlump-superextremal}
\end{center}
\end{figure}

One can further probe the near S-brane limit in the spirit of
Ref.~\cite{coordinates} by using a coordinate choice which
introduces a small hyperboloid near the S-brane worldvolume

\be \tau-\tau_+=T \frac{1+\cosh\tilde{\theta}}{2} , \hspace{.3in}
\sinh^2\theta = T \frac{\cosh\tilde{\theta} -1}{\tau_+ - M} \ee

\be d\tau = dT\frac{1+\cosh\tilde{\theta}}{2} +  T
\frac{\sinh\tilde{\theta}}{2} d\tilde{\theta} , \hspace{.3in}
d\theta = \frac{ dT (\cosh\tilde{\theta} -1) + T
\sinh\tilde{\theta} d\tilde{\theta}}{
2\sqrt{(\tau_+-M)T(\cosh\tilde{\theta}-1)}} \ee

\be \Delta= (T-\tau_+)(T-\tau_-) =  \tau_{+-} T
\frac{1+\cosh\tilde{\theta}}{2} \ . \ee

\noindent Using the twisted coordinate $\tilde{\phi}=\phi-Bz$ ,
the metric becomes

\begin{eqnarray}
ds^2_{twNE+} &=& \frac{M^2}{\tau_{+-}T/2}[-dT^2 +
T^2(d\tilde{\theta} + \sinh^2\tilde{\theta} d\tilde{\phi}^2) ] \nn \\
& & \ +\frac{T\tau_{+-}}{2\rho^2}[1+ \cosh\tilde{\theta} +
\frac{2}{\tau_{+-}^2/2} (a+B(\tau_+^2+a^2))^2 (\cosh\tilde{\theta}
-1) ]dz^2 \nn \\
& & \ +\frac{2T(\cosh\tilde{\theta}-1)}{2M^2 \tau_{+-}/2}
(\tau_+^2
+a^2)(a+B(\tau_+^2+a^2)) dz d\tilde{\phi} \nn \\
& & \ -\frac{T(\tau_+^2+a^2)^2}{4M^2 \tau_{+-}/2}
(\cosh\tilde{\theta}-1)^2 d\tilde{\phi}^2 \nn \\
&=& \ \frac{M^2}{\tau_{+-}T/2}[-dT^2 + T^2 dH_2^2]
-\frac{T}{\tau_{+-}/2} [ (\cosh\tilde{\theta}-1)d\tilde{\phi}
-\frac{1+2BM}{M} dz]^2 \nn  \\
&&+\
\frac{T\tau_{+-}}{4M^2}[1+\frac{2M^2(1+2BM)}{\tau_{+-}^2/2}](1+\cosh\tilde{\theta})
dz^2
\end{eqnarray}

\noindent which is similar to a near origin limit of a hyperbolic
space version of Taub-NUT (with an extra term).  It would be
interesting to find the exact solution corresponding to this
limit which would be a new type of S-brane.  This follows from the
fact that the isometry generated by the Killing vector along the
twisted direction has an isolated fixed point which is at
$\tau=\tau_+$ and $\theta=0$.  This new solution should be
interesting in the same way that ``nut'' solutions are canonical
examples of isolated fixed points for Euclidean spacetimes.
Apparently this new solution is not an analytic continuation of
Taub-NUT solutions but an exact solution is expected to be found
from a modified ansatz.

It would be interesting to further explore several properties of
this solution.  Although closed timelike curves do exist they
apparently are created only for a finite duration. Given the
special properties of the Kerr ergosphere, there could also be an
interesting physical significance of these S-brane ergosphere
regions.

\section{Discussion and Further Directions}

Spacelike branes play an interesting role in time dependent
physics.  In this paper we found smooth S-branes with reduced
R-symmetry and which have a spacetime twist which may interpreted
as a type of charge.  A feature of these solutions is that they
have extremal limits in which the S-brane worldvolume undergoes
an expanding de Sitter phase.  Although the lifetime and amount
of inflation of this phase are infinite, the configuration is
unstable and eventually disperses. Near (sub)extremal limits of
this solutions are also interesting and were shown to describe
the decay of a cone with a twisted circle at its tip.
Superextremal solutions produce closed timelike curves.

It is tempting to consider the twisted S-brane to be an example
of a closed string analogue of an inhomogeneous \cite{roll,
followS,inhom} open string tachyon solution some of which had
interpretations as describing the formation of lower dimensional
branes and strings.  For the rolling tachyon, the solutions with
energy equal to the the tachyon potential correspond to the
creation of a brane while solutions with less energy are smooth.
For the twisted S-brane when $a>M$ the solution is smooth while
the extremal limit, $a=M$, should then correspond to the
formation of an object.  Reducing the R-symmetry is effectively
analogous to introducing inhomogeneities in the rolling tachyon.

Several avenues are available to pursue such as generalizations
of these solutions, their embedding into string theory and
relationships to cosmology. Further details will appear in a
following paper.

\vspace{.3in}

{\bf Acknowledgements}

\noindent I would like to express my thanks for very enjoyable and
valuable discussions to Allan Adams, Chiang-Mei Chen, Greg Jones,
Joanna Karczmarek, Lubos Motl, Shinji Mukohyama, Joris
Raeymaekers, Koushik Ray, Prasanta Tripathy, Johannes Walcher,
Toby Wiseman and Mattias Wohlfarth, and I would especially like to
express my sincere gratitude to Pei-Ming Ho and Andrew Strominger.

This work is supported in part by the National Science Council,
the Center for Theoretical Physics at National Taiwan University,
the National Center for Theoretical Sciences, and the CosPA
project of the Ministry of Education.

\newcommand{\J}[4]{{\sl #1} {\bf #2} (#3) #4}
\newcommand{\andJ}[3]{{\bf #1} (#2) #3}
\newcommand{\AP}{Ann.~Phys.~(N.Y.)}
\newcommand{\MPL}{Mod.~Phys.~Lett.}
\newcommand{\NP}{Nucl.~Phys.}
\newcommand{\PL}{Phys.~Lett.}
\newcommand{\PR}{ Phys.~Rev.}
\newcommand{\PRL}{Phys.~Rev.~Lett.}
\newcommand{\PTP}{Prog.~Theor.~Phys.}
\newcommand{\hep}[1]{{\tt hep-th/{#1}}}


\begin{thebibliography}{10}

\bibitem{stro}
M.~Gutperle and A.~Strominger, {\sl ``Spacelike Branes,''}
\J{JHEP}{0204}{2002}{018},\\  {\tt hep-th/0202210}.

\bibitem{roll}
A.~Sen, ``Rolling Tachyon,'' \J{JHEP}{0204}{2002}{048}, {\tt
hep-th/0203211}; {\sl ``Tachyon Matter,''}
\J{JHEP}{0207}{2002}{065},  {\tt hep-th/0203265}; {\sl ``Field
Theory of Tachyon Matter,''} \J{\MPL}{A17}{2002}{1797}, {\tt
hep-th/0204143}; {\sl ``Time Evolution in Open String Theory,''}
\J{JHEP}{0210}{2002}{003}, {\tt hep-th/0207105}.

\bibitem{sugraSbranes}
\begin{description}
\item
C.~M.~Chen, D.~V.~Gal'tsov and M.~Gutperle, {\sl ``S-brane
solutions in supergravity theories,''} \J{\PR}{D66}{2002}{024043},
{\tt hep-th/0204071};
\item
M.~Kruczenski, R.~C.~Myers and A.~W.~Peet, {\sl ``Supergravity
S-branes,''} \J{JHEP}{0205}{2002}{039}, {\tt hep-th/0204144};
\item
S.~Roy, {\sl ``On supergravity solutions of space-like
Dp-branes,''} \J{JHEP}{0208}{2002}{025}, {\tt hep-th/0205198};
\item N.~S.~Deger and A.~Kaya, {\sl
``Intersecting S-brane solutions of D = 11 supergravity,''}
\J{JHEP}{0207}{2002}{038}, {\tt hep-th/0206057};
\item J.~E.~Wang, {\sl ``Spacelike and time dependent branes
from DBI,''} \J{JHEP}{0210}{2002}{037}, {\tt hep-th/0207089}.
\item
C.~P.~Burgess, F.~Quevedo, S.~J.~Rey, G.~Tasinato and C.~Zavala,
{\sl ``Cosmological spacetimes from negative tension brane
 backgrounds,''} \J{JHEP}{0210}{2002}{028},
{\tt hep-th/0207104};
\item
N.~S.~Deger, {\sl ``Non-standard intersections of S-branes in
D=11 supergravity,''} \J{JHEP}{0304}{2003}{034}, {\tt
hep-th/0303232} .
\end{description}




\bibitem{followS}
\begin{description}
\item O.~Obregon, H.~Quevedo and M.~P.~Ryan,
{\sl ``Kerr metric as an exact solution for unpolarized $S1\times
S_2$ Gowdy models,''} \J{\PR}{D65}{024022}{2002};
\item B.~McInnes,
{\sl ``Stringy instability of topologically non-trivial AdS black
holes and of de Sitter S-brane spacetimes ,''}
\J{\NP}{B660}{373}{2003}, {\tt hep-th/0205103};
\item K.~Ohta and T.~Yokono, {\sl ``Gravitational approach to
tachyon matter,''} \J{\PR}{D66}{2002}{125009}, {\tt
hep-th/0207004};
\item N.~Moeller
and B.~Zwiebach, {\sl ``Dynamics with infinitely many time
derivatives and rolling tachyons,''} \J{JHEP}{0210}{2002}{034},
{\tt hep-th/0207107};
\item
A.~Buchel, P.~Langfelder and J.~Walcher, {\sl ``On time-dependent
backgrounds in supergravity and string theory,''}
\J{\PR}{D67}{2003}{024011}, {\tt hep-th/0207214};
\item
A.~Buchel, P.~Langfelder and J.~Walcher, {\sl ``Does the tachyon
matter?,''} \J{Annals Phys.~}{302}{2002}{78}, {\tt
hep-th/0207235};
\item V.~D.~Ivashchuk, {\sl ``Composite S-brane
solutions related to Toda-type systems,''} \J{Class.~Quant.\
Grav.}{20}{2003}{261}, {\tt hep-th/0208101};
\item P.~Mukhopadhyay and A.~Sen, {\sl
``Decay of unstable D-branes with electric field,''}
\J{JHEP}{0211}{2002}{047}, {\tt hep-th/0208142};
\item T.~Okuda and S.~Sugimoto, {\sl ``Coupling
of rolling tachyon to closed strings,''}
\J{\NP}{B647}{2002}{101}, {\tt hep-th/0208196};
\item A.~Sen, {\sl ``Time and tachyon,''} \J{IJMP}{A18}{2003}{4869} , {\tt
hep-th/0209122};
\bibitem{Strotalk}
A.~Strominger, {\sl ``Open string creation by S-branes,''}
\textit{Cargese 2002, Progress in string, field and particle
theory} p.335, {\tt hep-th/0209090};
\item
B.~Chen, M.~Li and F.~L.~Lin, {\sl ``Gravitational radiation of
rolling tachyon,''} \J{JHEP}{0211}{2002}{050}, {\tt
hep-th/0209222};
\item
J.~Kluson, {\sl ``Exact solutions in open bosonic string field
theory and marginal  deformation in CFT,''}
\J{JHEP}{0312}{2003}{050},  {\tt hep-th/0209255};
\item
F.~Quevedo, G.~Tasinato and I.~Zavala, {\sl ``S-branes, negative
tension branes and cosmology,''} \textit{Hamburg 2002,
Supersymmetry and unification of fundamental interactions}
vol.~2, p. 1308, {\tt hep-th/0211031};
\item K.~Hashimoto, P.~-M.~Ho and J.~E.~Wang, {\sl
``S-brane Actions,''} \J{\PRL}{90}{2003}{141601}, {\tt
hep-th/0211090};
\item M.~Gutperle and A.~Strominger, {\sl
``Timelike boundary Liouville theory,''}
\J{\PR}{D67}{2003}{126002},  {\tt hep-th/0301038};
\item
N.~Ohta, {\sl ``Intersection rules for S-branes,''}
\J{\PL}{B558}{2003}{213}, {\tt hep-th/0301095};
\item C.~P.~Burgess, P.~Martineau, F.~Quevedo,
G.~Tasinato and I.~Zavala C., {\sl ``Instabilities and particle
production in S-brane geometries,''} \J{JHEP}{0303}{2003}{050},
{\tt hep-th/0301122};
\item A.~Maloney, A.~Strominger and X.~Yin, {\sl ``S-brane
thermodynamics,''} \J{JHEP}{0310}{2003}{048}, {\tt
hep-th/0302146};
\item
F.~Leblond and A.~W.~Peet, {\sl ``SD-brane gravity fields and
rolling tachyons,''} \J{JHEP}{0304}{2003}{048},  {\tt
hep-th/0303035};
\item
N.~Lambert, H.~Liu and J.~Maldacena, {\sl ``Closed strings from
decaying D-branes,''} {\tt hep-th/0303139};
\item
K.~Hashimoto, P.~M.~Ho, S.~Nagaoka and J.~E. Wang, {\sl ``Time
Evolution via S-branes,''} \J{\PR}{D68}{2003}{026007}, {\tt
hep-th/0303172};
\item
A.~Buchel and J.~Walcher, {\sl ``Comments on supergraivty
description of S-branes,''} \J{JHEP}{0305}{2003}{069}, {\tt
hep-th 0305055};
\item
F.~Leblond and A.~W.~Peet, {\sl ``A note on the singularity
theorem for supergravity SD-branes,''} {\tt hep-th/0305059};
\item
J.~L.~Karczmarek, H.~Liu, J.~Maldacena and A.~Strominger, {\sl
``UV finite brane decay,''} \J{JHEP}{0311}{2003}{042}, {\tt
hep-th/0306132};
\item
H.~L\"{u} and J.~F.~V\'{a}zquez-Poritz, {\sl ``Four-dimensional
Yang-Mills de Sitter gravity from eleven dimensions,''} {\tt
hep-th/0308104};
\item
H.~L\"{u} and J.~F.~V\'{a}zquez-Poritz, {\sl ``Smooth cosmologies
from M-theory,''} {\tt hep-th/0401150}.
\end{description}


\bibitem{nonsing}
G.~Jones, A.~Maloney and A.~Strominger, {\sl ``Non-singular
solutions for S-branes,''} {\tt hep-th/0403050}.

\bibitem{Sbranecosmo}
\begin{description}
\item
P.~K.~Townsend and M.~N.~Wohlfarth, {\sl ``Accelerating
cosmologies from compactification,''} \J{\PRL}{91}{2003}{061302},
{\tt hep-th/0303097};
\item
N.~Ohta, {\sl ``Accelerating cosmologies from S-branes,''}
\J{\PRL}{91}{2003}{061303}, {\tt hep-th/0303238};
\item
M.~N.~Wohlfarth, {\sl ``Accelerating cosmologies and a phase
transition in M-theory,''} Phys.\ Lett.\ B {\bf 563} (2003) 1,
{\tt hep-th/0304089};
\item
S.~Roy, {\sl ``Accelerating cosmologies from M/String theory
compactifications,''} Phys.\ Lett.\ B {\bf 567} (2003) 322, {\tt
hep-th/0304084};
\item
N. Ohta, {\sl ``A study of accelerating cosmologies from
Superstring/M theories,''} \J{\PTP}{110}{2003}{269},  {\tt
hep-th/0304172};
\item
R.~Emparan and J.~Garriga, {\sl ``A note on accelerating
cosmologies from compactifications and S-branes,''}
\J{JHEP}{0305}{2003}{028}, {\tt hep-th/0304124};
\item
C.~M.~Chen, P.~M.~Ho, I.~P.~Neupane and J.~E.~Wang, {\sl ``A note
on acceleration from product space compactification,''} \J{JHEP}
{0307}{2003}{017}, {\tt hep-th/0304177};
\item
M.~Gutperle, R.~Kallosh and A.~Linde, {\sl ``M / string theory,
S-branes and accelerating universe,''} \J{JCAP}{0307}{2003}{001},
{\tt hep-th/0304225};
\item
C.~M.~Chen, P.~M.~Ho, I.~P.~Neupane, N.~Ohta and J.~E.~Wang, {\sl
``Hyperbolic Space Cosmologies,''} \J{JHEP}{0310}{2003}{058},
{\tt hep-th/0306291};
\item
M.~Wohlfarth, {\sl ``Inflationary cosmologies from
compactification?,''} {\tt hep-th/0307179}.
\end{description}

\bibitem{imaginary}
\begin{description}
\item
A.~Maloney, A.~Strominger and X.~Yin, {\sl ``S-brane
thermodynamics,''} \J{JHEP}{0310}{2003}{048}, {\tt
hep-th/0302146};
\item
D.~Gaiotto, N.~Itzhaki and L.~Rastelli, {\sl ``Closed strings as
imaginary D-branes,''}  {\tt hep-th/0304192}.
\end{description}

\bibitem{Dowker}
\begin{description}
\item
F.~Dowker, J.~P.~Gauntlett, G.~W.~Gibbons and G.~T.~Horowitz,
{\sl ``The decay of magnetic fields in Kaluza-Klein theory,''}
 \J{\PR}{D54}{1995}{6929}, {\tt hep-th/9507143};
\item
F.~Dowker, J.~P.~Gauntlett, S.~B.~Giddings and G.~T.~Horowitz,
{\sl ``On pair creation of extremal black holes and Kaluza-Klein
monopoles,''} \J{\PR}{D50}{1994}{2662}, {\tt hep-th/9312172};
\item
F.~Dowker, J.~P.~Gauntlett, D.~A.~Kastor and J.~H.~Traschen, {\sl
``Pair creation of dilaton black holes,''}
\J{\PR}{D49}{1994}{2909}, {\tt hep-th/9309075 }.
\end{description}

\bibitem{closedtachyon}
\begin{description}
\item
A.~Adams, J.~Polchinski and E.~Silverstein, {\sl ``Don't Panic!
Closed string tachyons in ALE space-times,''}
\J{JHEP}{0110}{2001}{029}, {\tt hep-th/0108075};
\item
C.~Vafa, {\sl ``Mirror symmetry and closed string tachyon
condensation,''} {\tt hep-th/0111051};
\item
J.~R.~David, M.~Gutperle, M.~Headrick and S.~Minwalla, {\sl
``Closed string tachyon condensation on twisted circles,''}
\J{JHEP}{0202}{2002}{041}, {\tt hep-th/0111212}.
\end{description}

\bibitem{Hull}
\begin{description}
\item C.~Hull,
    {\sl ``Timelike T-duality, de Sitter space, large N gauge
    theories and topological field theory,''}
    \J{JHEP}{9807}{1998}{021}, {\tt hep-th/9806146};
    {\sl ``Duality and the signature of space-time,''}
    \J{JHEP}{9811}{1998}{017}, {\tt hep-th/9807127};
\item  V.~Balasubramanian, J.~de Boer and D.~Minic,
    {\sl ``Exploring de Sitter Space and Holography,''}
        \J{Class.~Quant.~Grav.}{19}{2002}{5655},
        {\tt hep-th/0207245};
\item  S.~Bhattacharya and S.~Roy,
    {\sl ``Time dependent supergravity solutions in arbitrary dimensions,''}
        \J{JHEP}{0312}{2003}{015},
        {\tt hep-th/030902}.
\end{description}

\bibitem{zeroduality}
\begin{description}
\item
M.~S.~Costa and M.~Gutperle, {\sl ``The Kaluza-Klein Melvin
solution in M theory,''} \J{JHEP}{0103}{2001}{027}, {\tt
hep-th/0012072};
\item
M.~Gutperle and A.~Strominger, {\sl ``Fluxbranes in string
theory,''} \J{JHEP}{0108}{2001}{037}, {\tt hep-th/0104136};
\item
O.~Bergman and M.~R.~Gaberdiel, {\sl ``Dualities of type 0
strings,''} \J{JHEP}{9907}{1999}{022}, {\tt hep-th/9906055}.
\end{description}

\bibitem{fbranelimit}
R.~Emparan and M.~Gutperle, {\sl ``From p-branes to fluxbranes
and back,''} \J{JHEP}{0112}{2001}{023}, {\tt hep-th/0111177}.

\bibitem{nullorb}
\begin{description}
\item
G.~T.~Horowitz and A.~R.~Steif, {\sl ``Singular string solutions
with nonsingular initial data,''} \J{\PL}{B258}{1991}{91};
\item
L.~Cornalba and M.~S.~Costa, {\sl ``A new cosmological scenario
in string theory,''} \J{\PR}{D66}{2002}{066001},  {\tt
hep-th/0203031};
\item
J.~Simon, {\sl ``The geometry of null rotation
identifications,''} \J{JHEP}{0206}{2002}{001},  {\tt
hep-th/0203201};
\item
H.~Liu, G.~Moore and N.~Seiberg, {\sl ``Strings in a
time-dependent orbifold,''} \J{JHEP}{0206}{2002}{045}, {\tt
hep-th/0204168};
\item H.~Liu, G.~Moore and N.~Seiberg, {\sl
``Strings in time-dependent orbifolds,''}
\J{JHEP}{0210}{2002}{031}, {\tt hep-th/0206182};
\item
L.~Cornalba and M.~S.~Costa, {\sl ``Time dependent orbifolds and
string cosmology,''} \J{Fortsch. Phys.}{52}{2004}{145}, {\tt
hep-th/0310099}.
\end{description}

\bibitem{coordinates}
A.~Sen, {\sl ``Strong coupling dynamics of branes from M
theory,''} \J{JHEP}{9710}{1997}{002}, {\tt hep-th/9708002}.

\bibitem{inhom}
\begin{description}
\item
F.~Larsen, A.~Naqvi and S.~Terashima, {\sl ``Rolling Tachyons and
Decaying Branes,''} \J{JHEP}{0302}{2003}{039},
 {\tt hep-th/0212248};
\item
A.~Ishida and S.~Uehara, {\sl ``Rolling down to D-brane and
tachyon matter,''} \J{JHEP}{0302}{2003}{050}, {\tt
hep-th/0301179};
\item
S.-J.~Rey and S.~Sugimoto, {\sl ``Rolling of Modulated Tachyon
with Gauge Flux and Emergent Fundamental String,''} {\tt
hep-th/0303133};
\item
A.~Fotopoulos and A.~A.~Tseytlin, {\sl ``On open superstring
partition function in inhomogeneous rolling tachyon
background,''} \J{JHEP}{0312}{2003}{025}, {\tt hep-th/0310253}.
\end{description}




\end{thebibliography}
\end{document}